# Visualizing Domain Ontology using Enhanced Anaphora Resolution Algorithm

L.Jegatha Deborah[1], R.Baskaran[1], A.Kannan[2]

[1] Department of Computer Science & Engineering, Anna University Chennai – 25
blessedjeny@gmail.com, baaski@annauniv.edu
[2] Department of Information Science & Technology, Chennai-25
kannan@annauniv.edu

**Abstract**

*Enormous explosion in the number of the World Wide Web pages occur every day and since the efficiency of most of the information processing systems is found to be less, the potential of the Internet applications is often underutilized. Efficient utilization of the web can be exploited when similar web pages are rigorously, exhaustively organized and clustered based on some domain knowledge (semantic-based) [1]. Ontology which is a formal representation of domain knowledge aids in such efficient utilization. The performance of almost all the semantic-based clustering techniques depends on the constructed ontology, describing the domain knowledge [6]. The proposed methodology provides an enhanced pronominal anaphora resolution, one of the key aspects of semantic analysis in Natural Language Processing for obtaining cross references [19] within a web page providing better ontology construction. The experimental data sets exhibits better efficiency of the proposed method compared to earlier traditional algorithms.*

## 1    Introduction

Semantic Analysis is a technique of relating syntactic structure inclusive of phrases, clauses, sentences or paragraphs. Bridging the semantic gap between heterogeneous systems is a prerequisite to information retrieval [7]. The basis of the above bridge is found in Ontology [1][2][5]. Ontology, in simple terms is a knowledge structure specifying the different terms and their relationships pertained to a particular domain [9-10]. Earlier systems performed semantic analysis with the help of ontology that consists of terms and relationships related to synonyms, antonyms, hyponyms, hypernyms and Thesaurus [3-6]. In the midst of such inventions, identifying and resolving the presence of anaphors and cataphora among the sentences pertained to a particular domain was a milestone to be achieved until 1998. Limiting our methodology to anaphora resolution, where the process of "Anaphora Resolution" (AR) or Pronouns resolution is the problem of resolving earlier reference of a phrase or a word in the same real-world entity and is found to be one of the complicated problems in Natural Language Processing [11-13]. There is a possibility that one sentence in a single domain can be referred from another sentence and such kind of relationships between sentences is called as co-referencing relationship. Coreferencing involves the detection of anaphor, where it refers to word or phrase in a sentence used to refer to an entity introduced earlier in the discourse [13] [19] . Resolving anaphora finds the best place in many of the applications including information extraction, information retrieval, NLP applications, semantic and web ontology. The three predominantly occurring types of anaphora are pronominal anaphora, definite noun phrase anaphora and one anaphora used in different application domains. The anaphora resolution process relies on some of the factors like gender, number agreement, semantic consistency, syntactic parallelism, proximity, etc [13]. Most of the traditional systems attempted to resolve anaphora in a single sentence. To be very specific, the anaphora resolution done by those systems was predominantly intra-sentential (the antecedent is present in the same sentence as that of anaphor) [14]. Compound words in the input corpus attempt to give meaningful information in anaphora resolution. The key strength of the enhanced pronominal anaphora resolution algorithm proposed in this paper provides inter-sentential anaphora resolutions by uncovering compound nouns and resolving the POS for each and every word. The proposed algorithm is found to work better on many web input text corpus as well as standard corpus provided by many universities as well. The experimental result of the proposed algorithm is compared with some of the traditional existing anaphora resolution methodologies which proved to have a better performance [15].

The remainder of this paper is organized as follows. Section 2 conducts a brief summary of the existing systems. Section 3 exhibits the system architecture and the working of the proposed algorithm. Section 4 illustrates the experimental results of the proposed algorithm with the comparison results shown. Section 5 presents the concluding remarks of the work.

## 2  Related Works

Hobbs' algorithm [16] relies on a simple tree search procedure formulated in terms of depth of embedding and left-right order. The tree procedure selects and replaces the pronouns by selecting the first candidate encountered by a left right depth first search for the tree. The algorithm chooses as the antecedent of a pronoun P the first $NP_i$ (Noun Phrase) in the tree obtained by left-to-right breadth-first traversal of the branches to the left of the path T. If an antecedent satisfying this condition is not found in the sentence containing P, the algorithm selects the first NP obtained by a left-to-right breadth first search of the surface structures of preceding sentences in the text. The algorithm is found to produce a success rate close to 80% for intrasentential anaphora resolution.

Shalom Lappin and Herbert Leass [17] report an algorithm for identifying the noun phrase antecedents of third person pronouns and lexical anaphors. The algorithm (hereafter referred to as RAP (Resolution of Anaphora Procedure) applies to the syntactic representations generated by McCord's Slot Grammar parser (McCord 1990, 1993) and relies on salience measures derived from syntactic structure and a simple dynamic model of attentional state to select the 12 antecedent noun phrase of a pronoun from a list of candidates . RAP algorithm concentrates more on resolving an intrasentential syntactic filter for ruling out anaphoric dependence of a pronoun on an NP on syntactic grounds. It employs an anaphor binding algorithm for identifying the possible antecedent binder of a lexical anaphor within the same sentence. The algorithm does not employ semantic conditions or real-world knowledge in choosing among the candidates. This algorithm is suited for intrasentential anaphora resolution, which will not be the case in most of the text corpus available in the WWW. RAP is also not suited in identifying the exact antecedents and replaces of such antecedents when the noun phrase is not a single but a compound noun phrase.  The major limitation of the algorithm is that the performance in terms of resolving the entire set of anaphor is found to be very limited when the input corpus consists of a number of compound noun phrases, even though the algorithm employs a decision procedure for selecting the preferred element of a list of antecedent candidates for a pronoun.

C. Aone and S. Bennet [18] describe an approach to building an automatically trainable anaphora resolution system. The authors made use of a machine learning algorithm and used many training examples for anaphora resolution.  This machine learning algorithm made use of a decision tree consisting of feature vectors for pairs of an anaphora and its possible antecedent. The feature vectors for the training samples include lexical, semantic, syntactic and positional features. The authors built 6 machine learning based anaphora resolvers and achieved about a precision close to 80%. However, the algorithm failed in cases when the machine learning algorithm has to resolve the anaphors between different sentences. The algorithm drastically showed lower performance when the intersentential anaphora resolution was performed.

Various Ontology construction techniques are available like TEXT-TO-ONTO Ontology Learning Environment [20], TextOntoEx [21], OntoLT[22]. Most of the available ontology construction methodologies fail in recovering the inter-sentential anaphors for refined ontology construction [8]. The proposed algorithm helps in the construction and visualization of ontology using the graphviz tool [27] as indicated in the algorithm below.

## 3   Enhanced Pronominal Anaphora Resolution Algorithm (KADE) – Proposed Algorithm

The motivation of our enhanced Pronominal Anaphora Resolution algorithm KADE was from the theoretical background provided in the previous work done by Poesio, M. and Mijail A. Kabadjov (2004), which was an attempt at providing the domain independent anaphora resolver. KADE follows the algorithmic steps similar to the algorithm given by the authors mentioned above with the exception that KADE resolves intersentential anaphors. The key power of KADE algorithm is that the existence of related anaphors found anywhere in the web input text corpus or standard corpus could be identified and replaced. Our proposed KADE algorithm, which is an enhancement of the previous one, is resolving the anaphors among the different sentences (intersentential anaphora detections). Increased efficiency in resolving the anaphors is obtained in this algorithm because the lexical knowledge with respect to a particular domain of the text corpus through Natural Language Processing is considered [5-6]. On performing many empirical tests on various input text corpus, the performance in retrieving the correct anaphors between different sentences (intersentential anaphors) was found to be better than many of the traditional works handled. Our proposed algorithm KADE however uses the output of Stanford Parser [23-24], but also found to work well on FDG parsers [25] and Charniak parsers too [26]. The overall architecture of the system is shown below

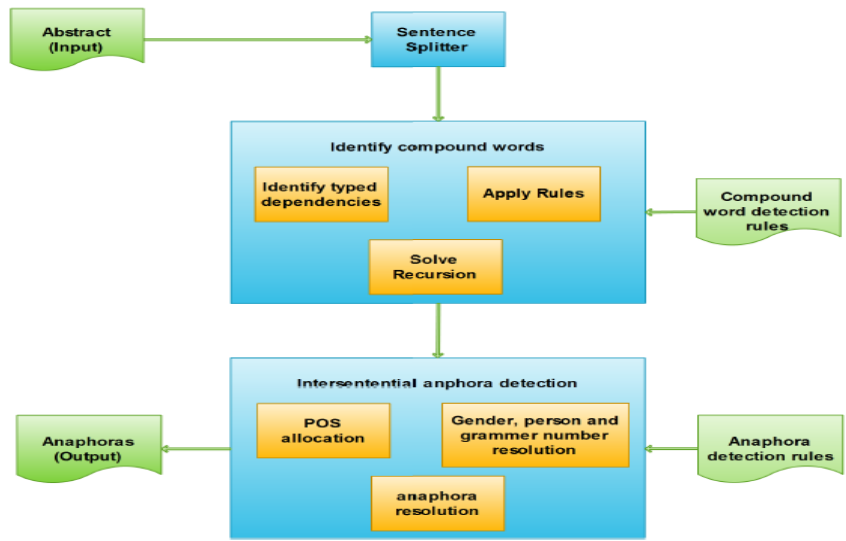

**Figure. 1.** System Architecture

The input to the algorithm is any type of web input text corpus (web search engines) of any length. Initially, all the sentences of the text corpus are provided with an identification number for the purpose of easy referencing. The typed dependencies among the different words in the raw sentences of text corpus are resolved using the traditional Stanford Parser [24]. Many unwanted dependencies may exist when using the parser and such dependencies must be removed. The cleaning process from the typed dependencies obtained earlier, is done by writing specific rules for identifying the compound words, lemmatizing the words and removing the unwanted tags. The compound words in our algorithm is identified by writing the rules like, a noun followed by another noun and a noun prefixed by adverbial modifiers is considered to be a compound noun. Once, the compound nouns are identified for the entire corpus, the document is cleaned by just deleting the unwanted tags. The anaphors existing in different sentences are identified by allocating an identifier like, CC for coordinating conjunction, DET for determiner, JJ for adjectives, NN for noun singular, NNS for noun plural, RB for adverb, etc and resolving the POS for each word in the sentence. Such identifier allocation and POS tagging is done using the Penn Tree bank [27]. On completion of the execution of identifying POS tagging for every word in the sentence, the list of anaphors are displayed. The algorithmic steps of the enhanced anaphora resolution algorithm KADE follow the procedure given below. The algorithm is initially tested against ordinary text corpus from the web. The enhanced algorithm can also be empirically tested against Brown Corpus (BC) and Susanne Corpus (SC), British National Corpus (BNC).

### 3.1 Algorithmic Procedure

Premise: Natural Language Processing

Domain: Text Corpus

*Input: Any web input text corpus*

*Output: List of Anaphors found*

Procedure:

Begin

do

{   // **Step 1**: Sentence Splitter

// **Step 2**: Resolving typed dependencies among the raw sentences

While (end of statement)

{   Assign Identifier Number for each sentence

     Describe the grammatical relationships in a sentence among words (nsubj, nn, det, prep, etc) }

// **Step 3**: Compound Nouns Identification using rules description

CW is set of Compound Words

    For each statement $S_i$ in S do

        $C_i \leftarrow$ Stanford Parser($S_i$)

        For each rule CWIRj in CWIR

            $CW_{new}$+=ApplyRule($CWIR_j$,$C_i$)

        i←i+1

    End

  For each statement $S_i$ in S do

    For each word $cw_{new}$ in CW

      If there exists cw=notDetected($cw_{new}$) in $S_i$

        $S_i \leftarrow$ Replace(cw,$cw_{new}$)

  End   End  return $C_i$  End

 // **Step 4**: Resolving Anaphors among the sentences (intersentential anaphora detections)

Begin

Do

{

Provide identifiers for each sentences, $S_1,S_2,…,S_n$

// Anaphora Resolution for he/she/it kind of words

If there exists word $W_i$ in sentence $S_k$ such that Personal-Pronoun($W_i$) is true then

  If there exists word $W_j$ in Sentence $S_{k-1}$ such that Noun($W_j$) is true then

    $W_i$ is anaphora of $W_j$

  Else  Display message "Unknown Phrase"

// Anaphora Resolution for who/where kind of words

    If there exists word $W_i$ in sentence $S_k$ such that POS($W_i$) ="WH" then

 If there exists word $W_j$ in Sentence $S_k$ such that Noun($W_j$) is true and Gender($W_i$)= Gender($W_j$) then $W_i$ refers to $W_j$

  Else Display message "Unknown Phrase"

    } While (end of document) End

Once the anaphors (Pronouns) are resolved, the ontology is constructed and visualized. The structural relationships that are obtained are made feasible through any visualization mechanism. For such visualization, a base ontology is used as a premise. The generated relationships are matched with the base ontology and create a new ontology graph. The base ontology is given as a graphViz dot file. The new relationships are matched with the base ontology and written back to graphViz dot file. The new ontology graph is now visualized using graphViz tool [27].

**Algorithm: Ontology Creation**

Input: Relationship file, Base ontology file

Output: Ontology graph

Procedure

Begin

Base ontology is given as graphViz dot file

Ontological Relationships file are mapped with the base ontology mapper file

New ontology is visualized using graphViz tool

End

## 4 Results and Discussions

The basic integrated development environment was developed to test the results for the experimental data sets done by KADE algorithm. The experimental tests were done on several raw input text corpuses. The performance efficiency in terms of correct retrieval of anaphors from the input corpus was found to be an average of 85%. Some of the sample data sets that were taken for the empirical tests for the exact retrieval of anaphors from the text corpus were Doctor Information System, Patient Information System, University Information System, Ontology Information Retrieval, etc [6]. The step by step screen shots for the algorithm evaluation are given below. Screen shots for a very small text corpus is shown

**Sample Input Text Corpus**

Every patient has a patient number. This number is used to identify the record of the patient. For every record there is a separate slot to hold the details of doctors who checked the patient and the medicines that they should take.

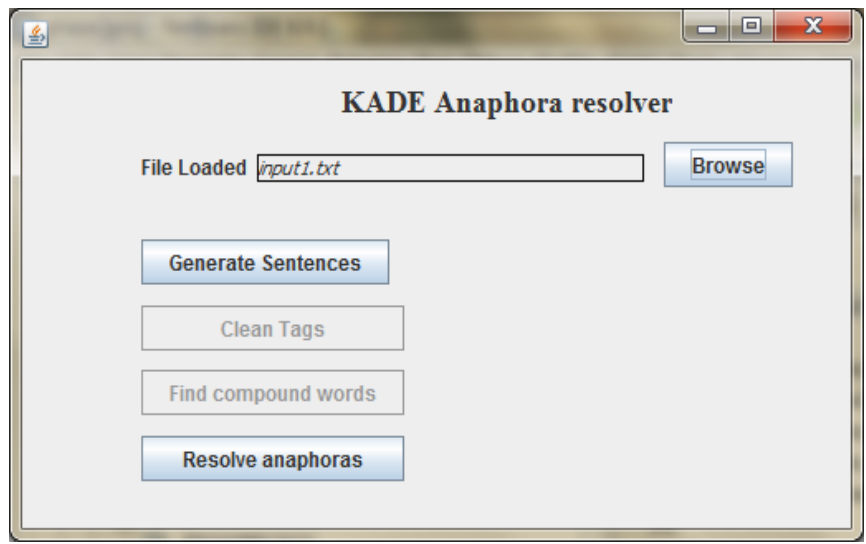

**Figure. 2.** Basic IDE for KADE

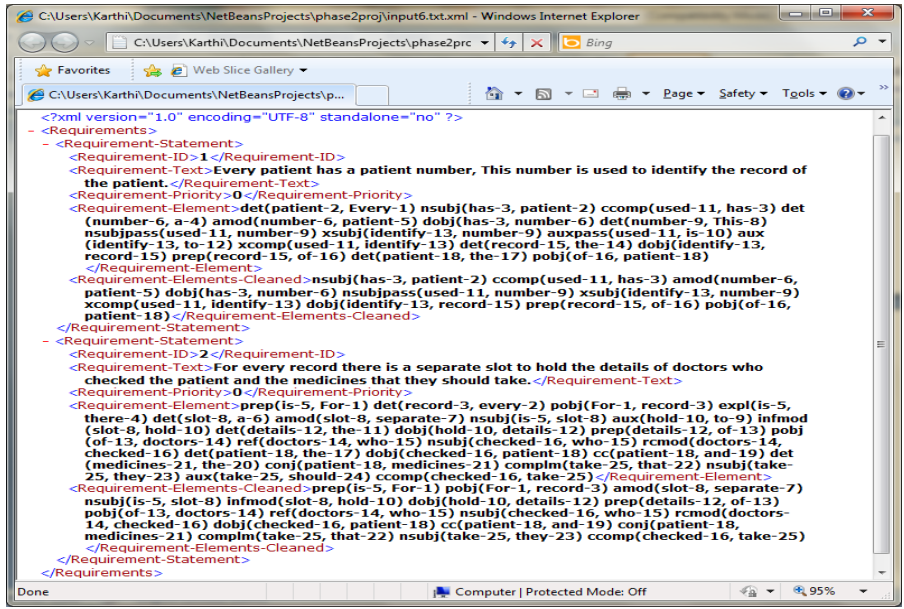

**Figure. 3.** Screen Shot of Parsed Text

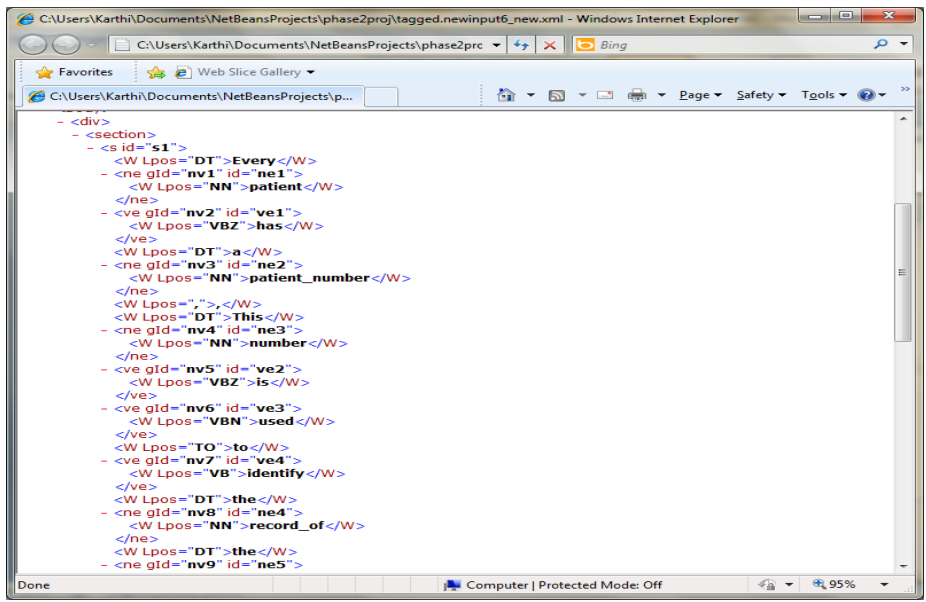

**Figure. 4.** Screen shot of POS tagging and tokenization

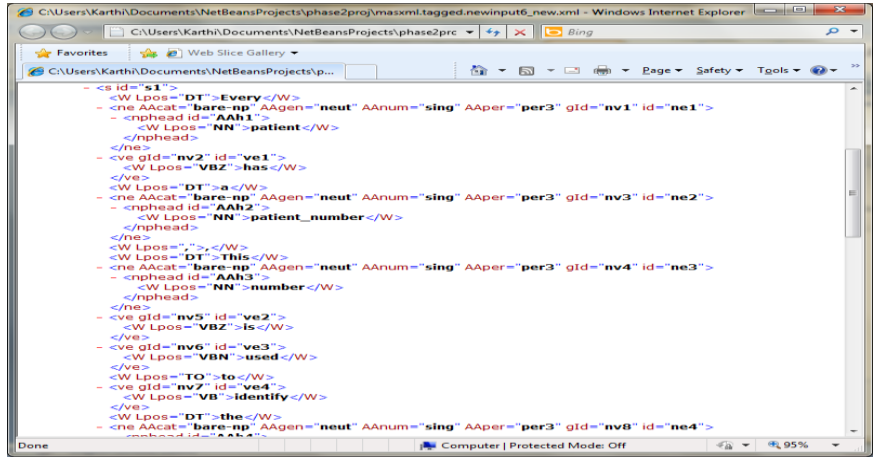

**Figure. 5.** Anaphors Resolution

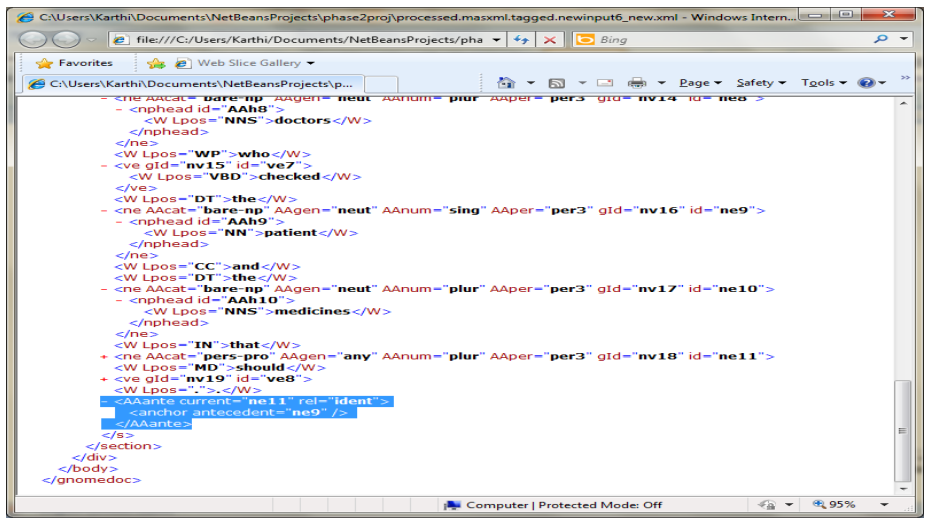

**Figure. 6.** List of Anaphors Resolved

The proposed KADE algorithm produces good formation of pronoun resolution. The following screen shots depict the scenario before and after anaphora resolution.

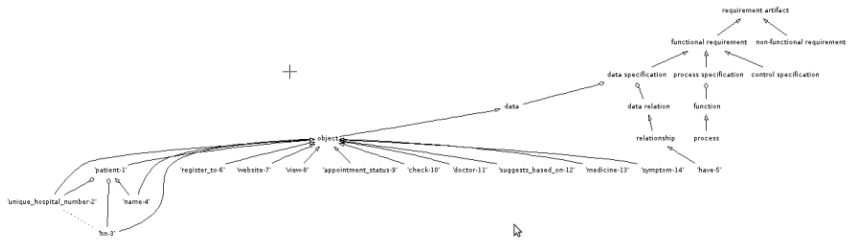

**Figure. 7** Before Anaphora Resolution

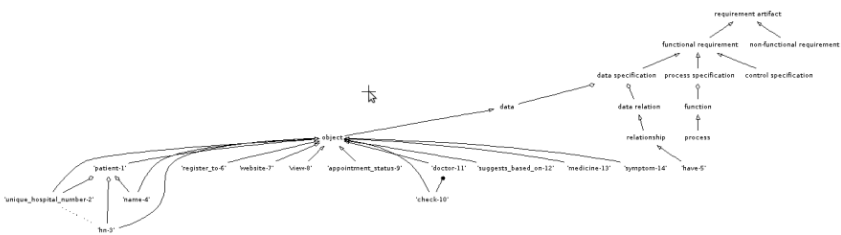

**Figure 8.** After Anaphora Resolution

The results of the KADE algorithm is compared with the other approaches and the graphical results below

## 4.1 Result Set 1:

KADE algorithm is compared with two approaches of Java RAP and MARS algorithms. The algorithms are compared for the total number of anaphors present against the total number of anaphors retrieved.

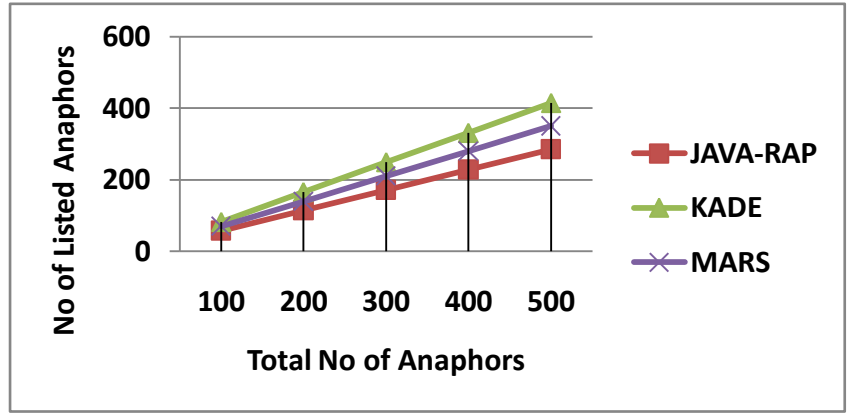

**Figure. 9.** Comparison Results

### 4.2 Result Set 2:

KADE algorithm is evaluated against the traditional performance parameters precision and recall. Precision evaluates the correct number of pronominal anaphors retrieved to the actual pronominal anaphors present in the corpus. Performance parameter recall evaluates the correct number of pronominal anaphors to the guessed pronominal anaphors in the corpus given by the domain expert. Precision and recall values are formulated as given below

**Assumption:**
Let k be the number of actual anaphors present in the text corpus.
Let c be the number of correct anaphors obtained from the text corpus using any anaphora resolution algorithm.
Let g be the number of correct anaphors given by the user, preferably a domain expert.

$$\text{Precision} = \frac{c}{k} \qquad (1)$$

$$\text{Recall} = \frac{c}{g} \qquad (2)$$

The experimental results for different data sets, randomly collected abstract documents from the web engines viz. Doctor Information System (DIS), Patient Information System (PIS), Ontology Information Retrieval (ORS), and their corresponding graphical results are shown below

**Table 1.** Evaluation Results - Precision

| Text Corpus Files | No. of Actual Anaphors | Precision Value | | |
|---|---|---|---|---|
| | | Hobb | Java RAP | KADE |
| Doctor Information System | 77 | 0.7 | 0.77 | 0.88 |
| Patient Information System | 57 | 0.5 | 0.7 | 0.84 |
| Ontology Information Retrieval | 130 | 0.82 | 0.8 | 0.9 |

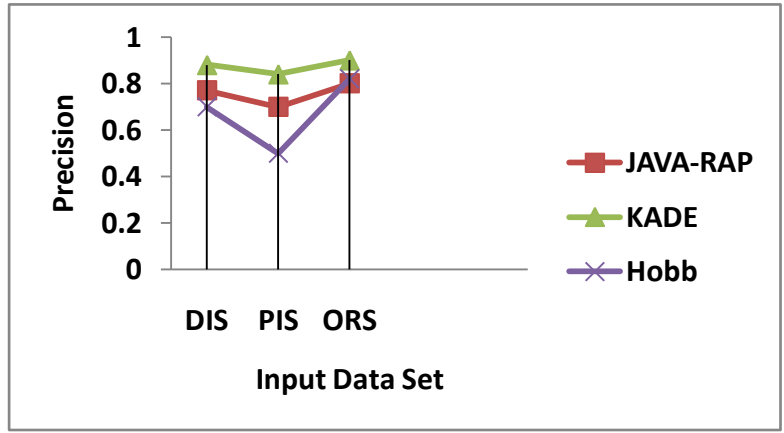

**Figure. 10.** Precision Comparison

**Table 2.** Evaluation Results – Recall

| Text Corpus Files | Number of Actual Anaphors | Recall Value | | |
|---|---|---|---|---|
| | | Hobb | Java RAP | KADE |
| Doctor Information System | 77 | 0.77 | 0.85 | 0.9 |
| Patient Information System | 57 | 0.68 | 0.8 | 0.88 |
| Ontology Information Retrieval | 130 | 0.85 | 0.90 | 0.96 |

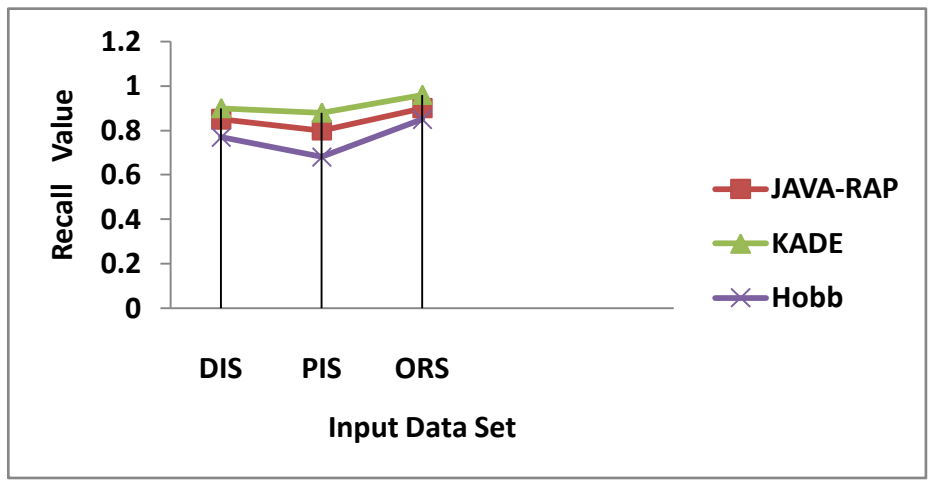

**Figure. 11. Recall Comparison**

## 5 Concluding Remarks

Ontology plays a vital role in clustering the web documents semantically to enhance the performance of many information extraction and information retrieval systems. Most of the systems given in the literature survey had the potential of constructing ontology based on synonyms, antonyms, hyponyms, anaphors and many more. This paper provides an enhanced pronominal anaphora resolution algorithm based on the results of Stanford Parser and Penn Treebank which works well on resolving anaphors existing among multiple sentences. The algorithm is tested against different data corpuses and is found to give better precision and recall values. The performance efficiency of the proposed algorithm in resolving intersentential anaphors is closer to 83%, compared to the traditional algorithms. This work provided a positive motivation and presents a wide research gap in the area of resolving cataphora in the raw text corpus which will be discussed in the future work.